\documentclass[pdftex,twocolumn,epjc3]{svjour3}          
\RequirePackage[T1]{fontenc}
\smartqed  
\RequirePackage{graphicx}
\RequirePackage{mathptmx}      
\RequirePackage{flushend}
\RequirePackage[numbers,sort&compress]{natbib}
\RequirePackage[colorlinks,citecolor=blue,urlcolor=blue,linkcolor=blue]{hyperref}

\journalname{Eur. Phys. J. C}
\usepackage{graphicx}
\usepackage{amssymb}
\usepackage{amsmath}
\usepackage{color}
\usepackage{hyperref}
 \usepackage{tabu}

\def\barray{\begin{array}}
\def\earray{\end{array}}
\def\be{\begin{equation}}
\def\ee{\end{equation}}
\def\ben{\begin{equation} \nonumber}
\def\een{\end{equation}}
\def\ban{\begin{eqnarray*}}
\def\ean{\end{eqnarray*}}
\def\ba{\begin{eqnarray}}
\def\ea{\end{eqnarray}}

\def\({\left(}
\def\){\right)}

\graphicspath{{./fig/}}

\begin{document}

\title{Tachyon Logamediate inflation on the brane}
\author{Vahid  Kamali\thanksref{e,addr}
,Elahe Navaee Nik \thanksref{e1,addr}}

\thankstext{e}{e-mail: vkamali@basu.ac.ir}
\thankstext{e1}{e-mail: e.navaeenik@gmail.com}
\institute{Department of Physics, Bu-Ali Sina University, Hamedan
65178, 016016, Iran\label{addr}
}
\date{Received: date / Accepted: date}
\maketitle
\begin{abstract}
According to a Barrow's solution for the scale factor of the universe, 
the main properties of the tachyon inflation model in the framework of RSII braneworld are studied.
Within this framework the basic slow-roll parameters are calculated analytically. 
We compare this inflationary scenario against 
the latest observational data. The predicted 
spectral index and the tensor-to-scalar fluctuation ratio 
are in excellent agreement with those of {\it Planck 2015}.
The current predictions are consistent 
with those of viable inflationary models.
\end{abstract}
\maketitle
\section{The set up and motivation:}
Standard model of inflation is driven by an scalar inflaton (quanta of the inflationary field) fields follow back to early
efforts to solve the basic problems of the Big-Bang cosmology, namely
horizon, flatness and  
monopoles \cite{Guth:1980zm, Albrecht:1982wi}.
The nominal inflationary paradigm contains two mainly different segments: the slow-roll and the 
(P)reheating regimes.
In the slow-roll phase the kinetic part of energy (which has the canonical form here) 
of the scalar field is negligible with respect to the potential part of energy $V(\phi)$
which implies a nearly deSitter expansion of the Universe.  
However, after the slow-roll epoch the kinetic energy 
becomes comparable to the potential energy and thus  
the inflaton field oscillates around the 
minimum at the (P)reheating phase and progressively the universe is filled by 
radiation \cite{Shtanov:1994ce,Kofman:1997yn}.
In order to achieve inflation one can use tachyon scalar fields
for which the kinetic term does not follow the canonical 
form (k-inflation \cite{ArmendarizPicon:1999rj}).
It has been found that tachyon fields which are associated with
unstable D-branes \cite{Sen:2002nu} may be responsible 
for the cosmic acceleration phase in early times 
\cite{Sen:2002an, Sami:2002fs, ArmendarizPicon:1999rj}. 
Notice, that tachyon potential has the following two properties:
the maximum of the potential occurs when
$\phi\rightarrow 0$ while the corresponding minimum takes place
when $\phi\rightarrow \infty$. For tachyonic models of inflation with ground state at $\phi\rightarrow \infty$, inflaton rolls toward its ground state without oscillating about it and the reheating mechanism does not work \cite{Kofman:2002rh}. For quasi power-low time dependence, which will be considered in the present work, there is a weak scale factor dependence of the tachyon energy density. Therefore in the post-inflation era the tachyon density would always dominate radiation unless there is a mechanism by which tachyon decay into radiation. Our tachyon model in the present work is an unphysical toy model but there is a solution for reheating problem in the context of warm inflation\cite{Setare:2012fg,Setare:2013ula,Setare:2014gya,Setare:2014uja,Setare:2013dd,Kamali:2016frd,Kamali:2017zgg,Basilakos:2017bol}, which however is beyond the scope of the
present work.
From the dynamical viewpoint one may present the equation of motion of tachyon field 
using a special Lagrangian 
\cite{Gibbons:2002md} which is non-minimally coupled to gravity:
\begin{equation}\label{Lag}
L=\sqrt{-g}\left[\frac{R}{16\pi G}-V(\phi)\sqrt{1-g^{\mu\nu}\partial_{\mu}\phi\partial_{\nu}\phi}\right]\;.
\end{equation}
Considering a spatially flat  
 $Friedmann-Lemaitre-Robertson-Walker$ (FLRW) (hereafter FLRW) universe 
the stress-energy tensor components are presented by 
\begin{equation}\label{1.1}
T^{\mu}_{\nu}=\frac{\partial
L}{\partial(\partial_{\mu}\phi)}\partial_{\nu}\phi-g^{\mu}_{\nu}L={\rm diag}(-\rho_{\phi},p_{\phi},p_{\phi},p_{\phi})
\end{equation}{equation}
where $\rho_{\phi}$ and $p_{\phi}$
are the energy density and pressure of the tachyon field. 
Combining the 
above set of equations one can find 
\begin{equation}\label{1.2}
\rho_{\phi}=\frac{V(\phi)}{\sqrt{1-\dot{\phi}^2}}
\end{equation}
and
\begin{equation}\label{1.22}
P_{\phi}=-V(\phi)\sqrt{1-\dot{\phi}^2}
\end{equation}
Where $\phi$ is tachyon scalar field in unite of inverse Planck mass $M_{pl}^{-1}$, and $V(\phi)$ is potential associated with the tachyon scalar field. 
In the past few years, there was a debate among particle physicists and 
cosmologists  regarding those 
phenomenological models which can be produced in extra dimensions.   
For example, the reduction of higher-dimensional 
gravitational scale, down to
TeV-scale, could be presented by an 
extra dimensional scenario \cite{ArkaniHamed:1998rs,ArkaniHamed:1998nn,Antoniadis:1998ig}. In these scenarios, gravity field
propagates in the bulk while 
standard models of particles are confined to the lower-dimensional brane.
In this framework, the extra dimension induces additional terms 
in the first Friedmann equation \cite{Binetruy:1999ut,Binetruy:1999hy,Shiromizu:1999wj}.
Especially, if we consider a quadratic term in the energy density 
then we can extract an accelerated expansion of the early universe \cite{Maartens:1999hf,Cline:1999ts,Csaki:1999jh,Ida:1999ui,Mohapatra:2000cm}. We will study tachyon inflation model in the framework of Randall-Sundrum II braneworld \cite{Randall:1999vf} which contains
a single, positive tension brane and a non-compact extra dimension. We note that this is not the only scenario in where these characteristics are presented. For example DBI Galileon inflation \cite{Silverstein:2003hf,Alishahiha:2004eh,Martin:2008xw,Guo:2008sz} has  these properties in its $T_3$ brane and cosmological inflation analysis of this model agrees with $1\sigma$ confidence level of Planck data \cite{Escamilla-Rivera:2015ova}.
Following the lines of Ref.\cite{Ravanpak:2015bta}, we attempt
to study the main properties of the tachyon inflation  
in which the scale factor evolves as $a(t)\propto \exp(A[\ln t]^{\lambda})$, where $\lambda>1$ and $A>0$
("Logamediate inflation"). For $\lambda=1$ case cosmic expansion evolves as ordinary power-law inflation $a\propto t^p$ where $p=A$ \cite{Barrow:1996bd,Barrow:2007zr}. More details regarding the cosmic expansion  
in various inflationary solutions can be found in the papers by
Barrow \cite{Barrow:1996bd,Barrow:2007zr}. In these papers there is no comment about the behaviour around $\lambda=0$ case of logamediate solution.  
In the current work, we investigate the possibility 
of using the logamediate solution in the case of 
 tachyon inflation on the brane. Specifically, the structure of 
the article is as follows: In section II we briefly
discuss the main properties of the 
tachyon inflation, while in section III we provide the perturbation  parameters.
In section IV we study the
performance of our predictions 
against the \textit{Planck 2015} data.
Finally, the main conclusions are
presented in section VI. 
\section{Tachyon inflation}
 In this part we consider a FLRW universe  with tachyon  component in inflation era, the basic cosmological equations in the context 
of the Randall-Sundrum II (RSII)  
brane \cite{Randall:1999vf} are presented as:
\begin{equation}\label{2.3}
H^2=\frac{1}{3}\rho_{\phi}(1+\frac{\rho_{\phi}}{2\tau}),
\end{equation}

\begin{equation}\label{denn}
{\dot \rho}_{\phi}+3H(\rho_{\phi}+p_{\phi})=0
\end{equation}
and
\begin{equation}\label{denn1}
\dot{H}=-\frac{1}{2M_p^2}(\rho_{\phi}+P_{\phi})(1+\frac{\rho_{\phi}}{\tau})
\end{equation}

where $H=\frac{\dot{a}}{a}$ and $a$ are Hubble parameter and scale factor respectively, dot means derivative with respect to the cosmic time. Parameter $\tau$ in Eq.(\ref{2.3}) represents the brane tension \cite{Shiromizu:1999wj, Binetruy:1999ut, Binetruy:1999hy}. The value of this term is constrained, to be $\tau>(1MeV)^4$ by considering   nucleosynthesis epoch \cite{Cline:1999ts}. Another stronger limitation for the value of $\tau$ is presented by usual tests for deviation from Newton's law $\tau\geq (10 Tev)^{4} $ \cite{Brax:2003fv}.  
The model is considered in natural unit $8\pi G=\frac{h}{2\pi}=c=1$. Using Eqs.(\ref{1.2},\ref{1.22},\ref{2.3},\ref{denn},\ref{denn1}), we can derive background evolution motions  of tachyon scalar field coupled by scale factor in high energy limit $\rho_{\phi}\gg\tau$.  
\begin{eqnarray}\label{E.O.M}
3H^2=\frac{V^2(\phi)}{2\tau\sqrt{1-\dot{\phi}^2}}~~~~~~~~~\\
\nonumber
\frac{\ddot{\phi}}{1-\dot{\phi}^2}+3H\dot{\phi}+\frac{1}{V}\frac{d V}{d\phi}=
0 \;,
\\
\nonumber
\dot{H}=-\frac{\rho^2\dot{\phi}^2}{2M_p^2\tau}~~~~~~~~~~~~~~~
\end{eqnarray} 
In Ref.\cite{Barrow:2007zr}, a complete analysis around the slow-roll parameters was made for canonical scalar fields which leads to slow-roll condition $3H\dot{\phi}\simeq -\frac{dV(\phi)}{d\phi}$. We will consider our model in slow-roll limit of tachyonic scalar fields ,$\dot{\phi}\ll 1$ ~~$\ddot{\phi}\ll 3H\dot{\phi}$  which leads to  $3HV\dot{\phi}\simeq -\frac{dV(\phi)}{d\phi}$ \cite{Fairbairn:2002yp}.  
In the slow-roll regime of tachyon fields, from relations (\ref{E.O.M}) $\dot{\phi}$ can be presented in term of Hubble parameter and its derivative:
\begin{eqnarray}\label{8}
\dot{\phi}=\sqrt{-\frac{\dot{H}}{3H^2}}
\end{eqnarray}
which will be used in our future purposes. Using Eqs.(\ref{denn1}) and (\ref{E.O.M}) we can find a real velocity field $\dot{\phi}$ in Eq.(\ref{8}). 
 Now we consider logamediate inflation model in which its scale factor behaves as \cite{Barrow:1996bd,Barrow:2007zr}:  
\begin{eqnarray}
a(t)=a_0\exp(A[\ln t]^{\lambda})\;,
\end{eqnarray}
one can present the compact solution of Eq.(\ref{8}).
 \begin{eqnarray}
\frac{d\phi}{dt}\simeq\sqrt{\frac{1}{A\lambda(\ln t)^{\lambda-1}}},\Rightarrow \phi-\phi_0=\frac{1}{\sqrt{A\lambda}}\int(\ln t)^{\frac{1-\lambda}{2}}dt
\end{eqnarray}
which leads to
\begin{eqnarray}
\phi=\phi_0+\frac{g(t)}{K}~~~~~~~~~~~\\
\nonumber
g(t)=\gamma(\frac{3-\lambda}{2},-\ln t)
\end{eqnarray} 
where $K=\sqrt{3\lambda A}$ and $\gamma(a,x)$ is incomplete gamma function \cite{Arfken,Abramowitz}, where $a$ is  an integer constant and $x$ is a variable,  for example in our case $(a,x)=(\frac{3-\lambda}{2},-\ln t)$ .  
Dimensionless slow-roll parameters of the model can be introduced by standard definition in term of scalar field 
\begin{eqnarray}
\epsilon=-\frac{\dot{H}}{H^2}=\frac{[\ln (g^{-1}[K(\phi-\phi_0)])]^{1-\lambda}}{\lambda A}~~~~~~~~~~~~~~~~~~\\
\nonumber
\eta=\frac{1}{2H}[-\frac{\ddot{V}}{\dot{V}}+\frac{\dot{H}}{H}+\frac{\dot{V}}{V}]=\epsilon[-1+\frac{1}{g^{-1}[K(\phi-\phi_0)]}]
\end{eqnarray}
where $g^{-1}(t)$ is inverse function of $g(t)$. 
In the above relations we have used the approximation, $\ln t\gg \lambda-1$ which may be used at the early time. Number of e-folding can be presented for the model 
\begin{eqnarray}\label{N}
N=\int_{t_1}^{t} Hdt'=\int_{\phi_1}^{\phi}\frac{H}{\dot{\phi'}}d\phi' ~~~~~~~~~~~~~~~~\\
\nonumber
N=(\ln[g^{-1}(K[\phi-\phi_0])])^{\lambda}-(\lambda A)^{\frac{\lambda}{1-\lambda}}
\end{eqnarray}
where $\phi_1$ is introduced at the beginning of the inflation when the $\epsilon =1$. Using Eq.(\ref{N}) we can find tachyon scalar field in term of variable number of e-fold $N$
\begin{eqnarray}
\phi=\phi_0+g[\exp([N+(\lambda A)^{\frac{\lambda}{1-\lambda}}]^{\frac{1}{\lambda}})]
\end{eqnarray}
which will be used for the future goals. Potential of tachyon filed may be presented by using Eq.(\ref{E.O.M}) 
\begin{eqnarray}
V(\phi)=6\tau (\lambda A)^2\frac{(\ln g^{-1}[K(\phi-\phi_0)])^{2(\lambda-1)}}{(g^{-1}[K(\phi-\phi_0)])^2}
\end{eqnarray}  
\section{Perturbation}
Although assuming a spatially-flat, isotropic and homogeneous FRW universe may be useful and reasonable, but there are observed deviations from isotropic and homogeneity in our universe. These deviations motivate us to use perturbation theory in cosmology. In the context of general relativity and gravitation, inhomogeneity grows with time, so it was very small in the past time. Therefore first order or linear perturbation theory can be used for scalar field models at the inflation epoch. Considering Einstein's equation, inflaton field in the FRW universe connects to the metric components of this universe, so Perturbed inflaton field must be studied in the perturbed FRW geometry. Most general linear perturbation of spatially-flat FRW metric is presented by:
\begin{eqnarray}
ds^2=-(1+2C)dt^2+2a(t)D_{,i}dx^i dt~~~~~~~~~~~~~\\
\nonumber
+a^2(t)[(1-2\psi)\delta_{ij}+2E_{,ij}+2h_{ij}]dx^{i}dx^{j}
\end{eqnarray}     
which includes scalar perturbations $C, D, \psi, E$ and traceless-transverse tensor perturbations $h_{ij}$. Power-spectrum of the curvature perturbation $P_R$, that is derived from correlation of first order scalar field perturbation in vacuum state can be constrained by observational data. For tachyon fields, this parameter at the first level is presented by \cite{Hwang:2002fp,Ravanpak:2015bta}
\begin{eqnarray}\label{P}
\mathcal{P}_R=(\frac{H^2}{2\pi\dot{\phi}})^2\frac{1}{V(1-\dot{\phi}^2)^{\frac{3}{2}}}
\end{eqnarray}  
This parameter is essential for our perturbed analysis which is presented in Ref.\cite{Hwang:2002fp}.
In slow-roll and high energy limit, using Eq.(\ref{E.O.M}), we may simplify the above relation as:
\begin{eqnarray}
\mathcal{P}_{R}=\frac{1}{4\pi^2 V}(\frac{V^2}{2\tau V'})^2~~~~~~~~~~~~~~~~~~~~~~
\\
\nonumber
\simeq\frac{3(A\lambda)^6}{4\pi^2\sqrt{6\tau}}\exp(-(\frac{N}{A})^{\frac{1}{\lambda}})(\frac{N}{A})^{\frac{4(\lambda-1)}{\lambda}}
\end{eqnarray}
where $V'=\frac{dV}{d\phi}$. 
Another two important perturbation parameters are spectral index $n_s=1+\frac{d\ln P_R}{d\ln k}$ and its running $n_{run}=\frac{d n_s}{d\ln k}$. From Eq.(\ref{P}) in the slow-roll limit, these parameters are presented by
\begin{eqnarray}
n_s=1+2\epsilon[-2+\frac{1}{g^{-1}(K[\phi -\phi_0])}]\simeq 1+\frac{4(1-\lambda)}{\lambda}\frac{1}{N}\\
\nonumber
n_{run}=\frac{4(\lambda-1)}{(\lambda A)^2}(\ln g^{-1}[K(\phi-\phi_0)])^{1-2\lambda}~~~~~~~~~~~~~~~~~~
\end{eqnarray}  
These parameters also may be constrained by observational data. Up to now, we consider scalar perturbation parameters. During inflation era, there are two independent components of gravitational waves $h_{+}, h_{\times}$ or tensor perturbation of the metric with the same equation of motion. Amplitude of the tensor perturbation is given by
\begin{eqnarray}
\mathcal{P}_{g}=8(\frac{H}{2\pi})^2(\frac{3}{\tau^2})^{\frac{1}{4}}H^{\frac{1}{2}}=\frac{2}{\pi^2}(\frac{3}{6^5\tau^7})^{\frac{1}{4}}V^{\frac{5}{2}}
\end{eqnarray}  
which have been presented in Ref.\cite{Langlois:2000ns}. 
Tensor-scalar ratio is another important parameter
\begin{eqnarray}
r=\frac{\mathcal{P}_g}{\mathcal{P}_R}=(\frac{2\pi^8 4^3}{3^4\tau^7})^{\frac{1}{4}}\frac{(2\tau V')^4}{V^{\frac{9}{2}}}~~~~~~~~~~~~~~~~~~~~~~~~~~~~~\\
\nonumber
\simeq\frac{197}{K}(\frac{N}{A})^{\frac{\lambda-1}{2\lambda}}\exp(-\frac{3}{2}(\frac{N}{A})^{\frac{1}{\lambda}})~~~~~~~~~~~~~~~~~~~~~~~~~~~~~\\
\nonumber
=\frac{197}{K}(\frac{4(\lambda-1)}{a\lambda}\frac{1}{1-n_s})^{\frac{\lambda-1}{2\lambda}}\exp(-\frac{3}{2}(\frac{4(\lambda-1)}{A\lambda(1-n_s)})^{\frac{1}{\lambda}})
\end{eqnarray}  
$r-n_s,$ trajectory for inflation models can be compared with Planck observational data.
\section{Comparison with observation}
\begin{figure}
\begin{center}
\includegraphics[scale=0.47]{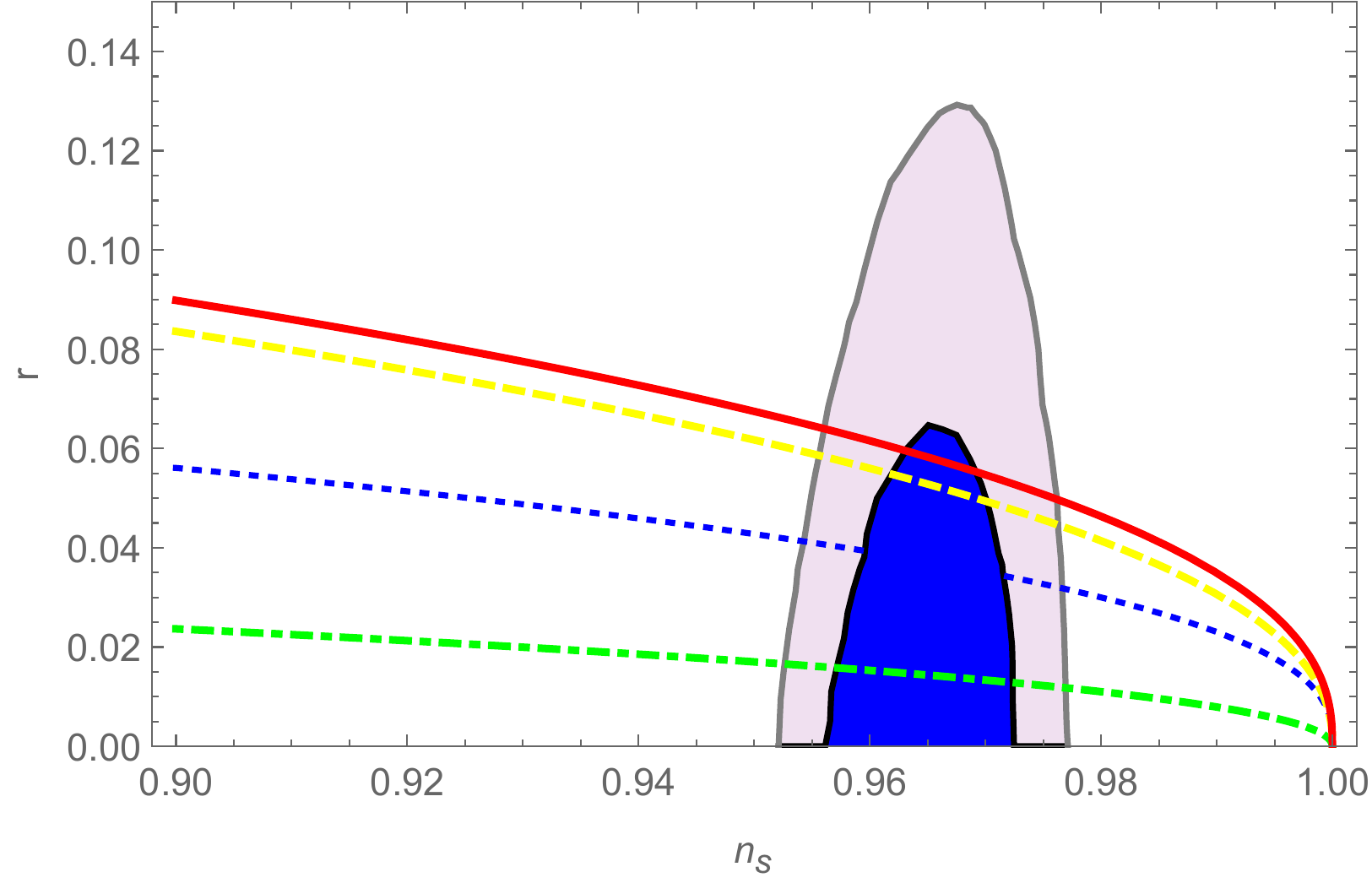}
\end{center}
\caption{$1\sigma$ and $2\sigma$ confidence regions which borrowed from Planck \cite{Planck:2013jfk},  $r-n_s$ trajectories of the present model. The solid red, dashed yellow, dotted  blue and dot-ashed green lines correspond to combinations $(29,4\times 10^{-12})$, $(39,10^{-15})$, $(19,3\times 10^{-6})$ and $(49, 5\times 10^{-4})$ of $(\lambda, A)$. There is no transition from $n_s<1$ to $n_s>1$.     }
\label{fig1}
\end{figure}

\begin{figure}
\begin{center}
\includegraphics[scale=0.47]{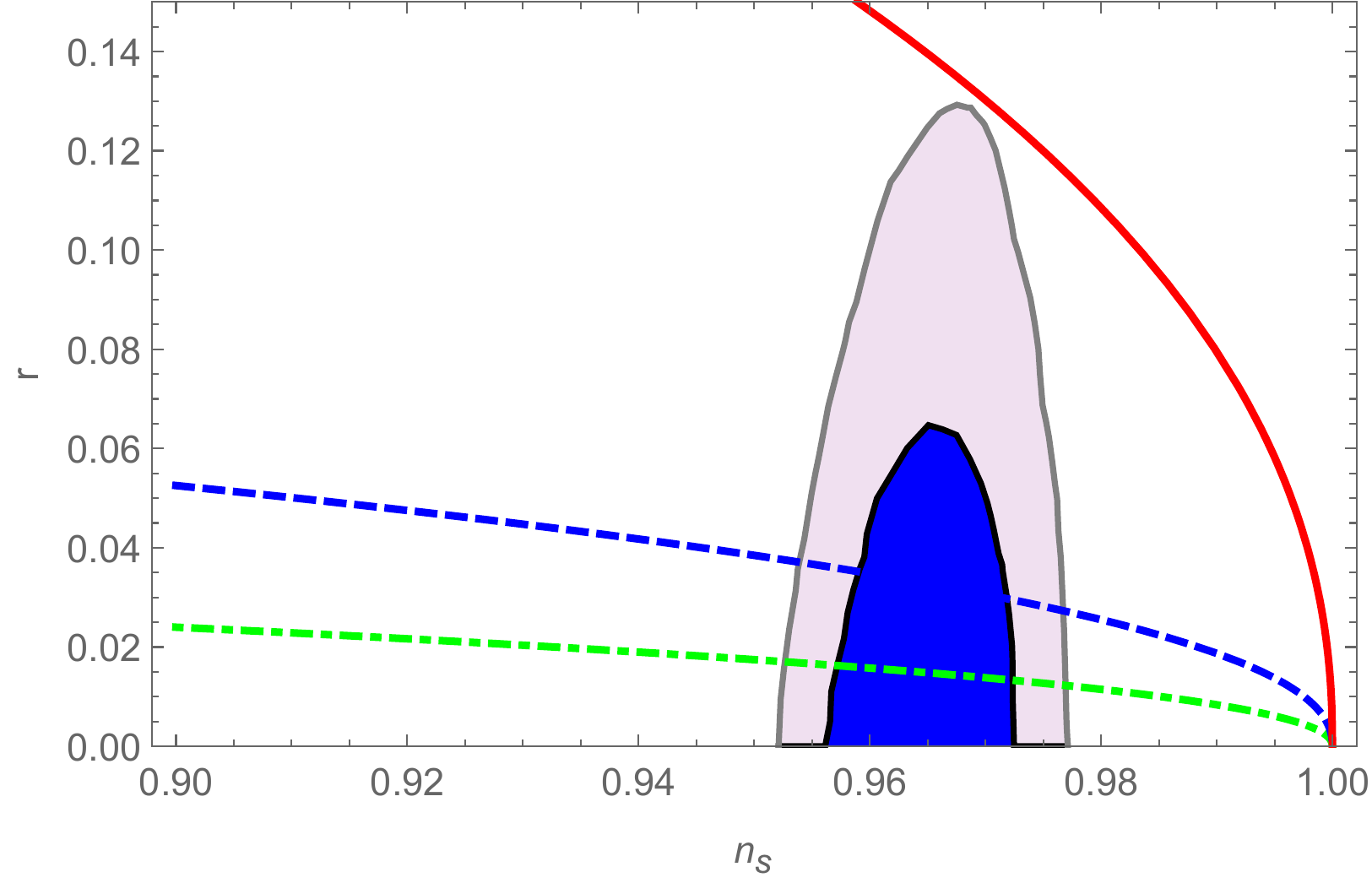}
\end{center}
\caption{$1\sigma$ and $2\sigma$ confidence regions which borrowed from Planck \cite{Planck:2013jfk},  $r-n_s$ trajectories of the present model. The dot-dashed green, dashed blue  and solid red lines related to pairs $(29, 12\times 10^{-3}),(3,10^{-12})$ and $(69,10^{-50})$. The trajectory placed out of $95\%$ confidence for the large value of $\lambda$.    }
\label{fig2}
\end{figure}

The analysis of \textit{Planck} data sets has been done in Ref.\cite{Ade:2015lrj}. The results of this analysis indicate the single scalar field models of inflation in slow-roll limit have limited spectral index, very low spectral running and tensor-scalar ratio.
\begin{eqnarray}\label{lim}
n_s=0.968\pm 0.006 ~~~~~~~~~~~~~~~~~~\\
\nonumber
r=\frac{\mathcal{P}_g}{\mathcal{P}_R}< 0.11,  ~~~~~~~~~~~~~~~~~~~~~~~\\
\nonumber
n_{run}=\frac{dn_s}{d\ln k}=-0.003\pm 0.007~~~~~
\end{eqnarray}   
The upper bound set on tensor-scalar ratio function and running of the tensor-scalar ratio has been gained by using the results of Planck team and joint analysis of BICEP2/Keck Array/Planck \cite{joint}. In the present section we will try to test the performance of tachyon inflation against the results of observation (\ref{lim}).
In Fig.(\ref{fig1}) we render the confidence contours in the $(n_s,r)$ plane. The values of pair $(\lambda,A)$ are fixed for each trajectory. The curves are related to the pairs $(\lambda,A)$ as: $(29,4\times 10^{-12})$, $(39,10^{-15})$, $(19,3\times 10^{-6})$ and $(49, 5\times 10^{-4})$ up to down. The main difference between our braneworld model and ordinary scalar field models \cite{Barrow:1996bd,Ravanpak:2015bta} is that there is no transition from $n_s<1$ to $n_s>1$ for all values of $\lambda$. For the big values of $\lambda$ with special combinations of $(\lambda, A)$ there are curves which behave as Harrison-zel'dovich spectrum i.e. $n_s=1$.
 In Fig.(\ref{fig2}), the dot-dashed green line and dashed blue line are related to the pairs $(29, 12\times 10^{-3}),(3,10^{-12})$ respectively. In this figure for the large value of $\lambda=69$ and small value of $A=10^{-50}$, the trajectory placed out of $95\%$ confidence which means large values of $\lambda$ are not compatible with Planck data.
   In Figs.(\ref{fig1}) and (\ref{fig2}) the curves of our model compared with $68\%$ and $95\%$ confidence regions from Planck 2015 result\cite{Ade:2015lrj} at $k_{*}=0.05 Mpc^{-1}$. 
\begin{figure}
\begin{center}
\includegraphics[scale=0.47]{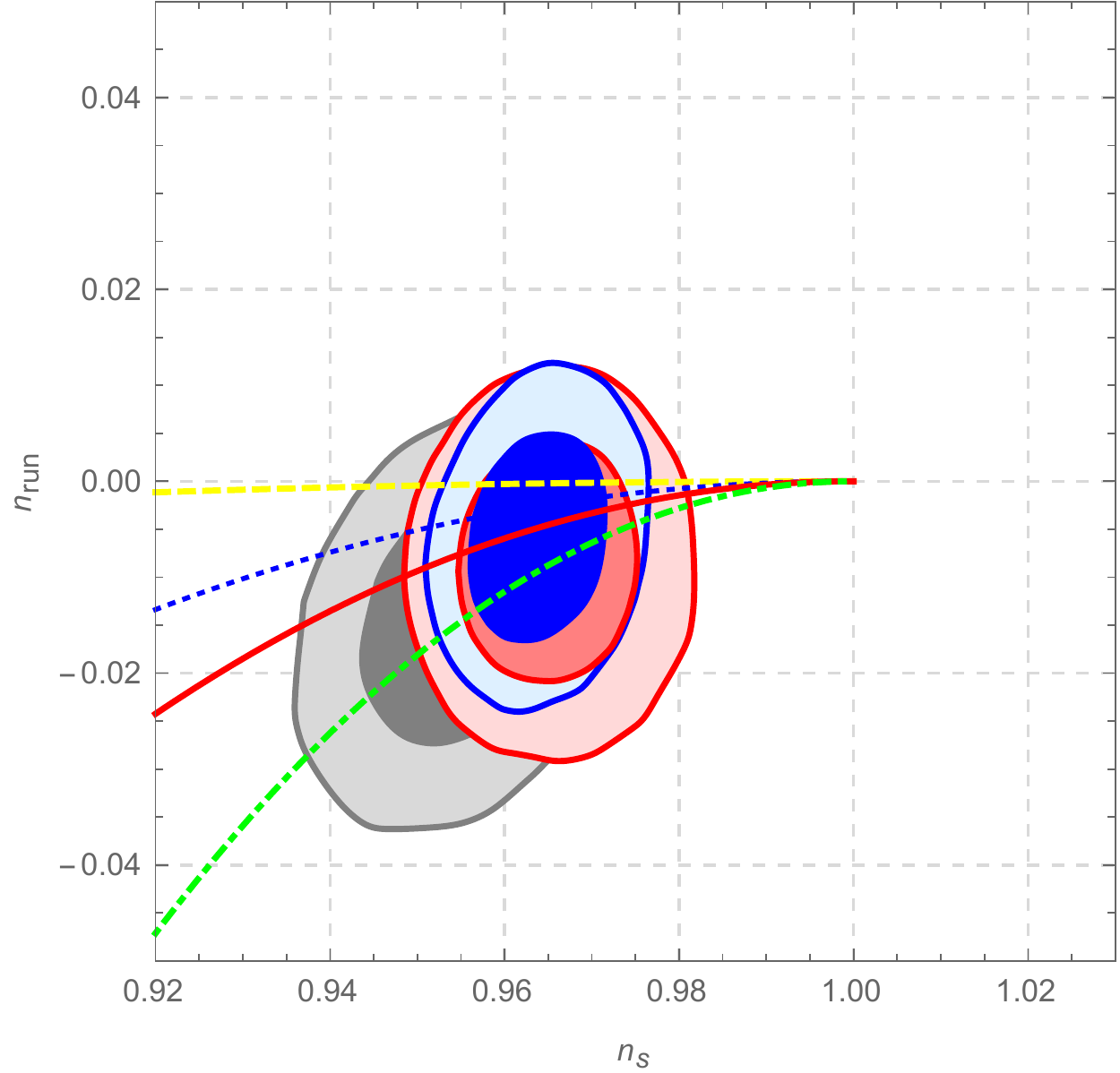}
\end{center}
\caption{ $n_s-n_{run}$ diagram. The area corresponds to Planck data and $n_s-n_{run}$ trajectories relate to our model. The dashed yellow, dotted blue, solid red, and dot-dashed green lines correspond to combinations: $(19,6\times 10^{-9})$, $(29, 4\times 10^{-3})$, $(39,2\times 10^{-20})$, $(49, 3\times 10^{-9})$}
\label{fig3}
\end{figure} 
In Fig.(\ref{fig3}), we plotted $n_{run}-n_{s}$ trajectories for some pairs $(\lambda, A)$  
which have been used in previous figures. There is no running in the scalar spectral index for combination $(19,6\times 10^{-9})$.
\section{Comparison with other models} 
Bellow we will compare the current predictions with those 
of viable literature potentials. 
This can help us to understand the variants  
of the tachyon-brane inflationary model
from the observationally viable inflationary scenarios.
\begin{itemize}
\item The Starobinsky or $R^{2}$ inflation model \cite{staro}:
In Starobinsky inflation model the asymptotic behavior 
of the effective potential is presented as 
$V(\phi)\propto\lbrack1-2\mathrm{e}%
^{-B\phi/M_{pl}}+\mathcal{O}(\mathrm{e}^{-2B\phi/M_{pl}})]$
which provides the following  predictions in the slow-roll limit\cite{Muk81,Ellis13}:
$r\approx 8/B^{2}N^{2}$ and ,$n_{s}\approx 1-2/N$  where $B^{2}=2/3$.
Therefore, if we select $N=50$ then we obtain
$(n_{s},r)\approx(0.96,0.0048)$. For $N=60$ we
find $(n_{s},r)\approx(0.967,0.0033)$. 
It has been found that the Planck data \cite{Ade:2015lrj}
favors the Starobinsky inflation. Obviously, our results 
(see figures \ref{fig1} and \ref{fig2}) are consistent with those of $R^{2}$ inflation.
\item The chaotic model of inflation \cite{Linde}:
In this inflationary model the potential is given by
$V(\phi) \propto \phi^{k}$. 
The basic slow-roll
parameters for this potential are presented as $\epsilon=k/4N$, $\eta=(k-1)/2N$
which leads to $n_{s}=1-(k+2)/2N$ and $r=4k/N$. 
It has been found that 
monomial potentials with $k\ge 2$ are not  in agreement with the Planck
data \cite{Ade:2015lrj}. 
Using $k=2$ and $N=50$ we present  
$n_{s}\simeq 0.96$ and $r\simeq 0.16$. For 
$N=60$ we find $n_{s}\simeq 0.967$ and $r\simeq 0.133$. 
It is interesting to note that 
this model also
corresponds to the results of intermediate inflation \cite{Barrow:1990vx,Barrow:1993zq,Barrow:2006dh,Barrow:2014fsa} with
Hubble rate during inflation which is given by $H\propto t^{k/(4-k)}$ with
$n_{s}=1-(k+2)r/8k$ and $k=-2$ case gives $n_{s}=1$ exactly to the first order. 
\item Hyperbolic model of inflation \cite{Basilakos:2015sza}:
In hyperbolic inflation the potential is
presented by $V(\phi) \propto \mathrm{sinh}^{b}(\phi/f_{1})$.
Initially,
 this potential was proposed in the context of late time acceleration phase or dark energy  \cite{Rubano:2001xi}.
Recently,  the properties
of this scalar field potential have been investigated back in the inflationary epoch \cite{Basilakos:2015sza} .
The slow-roll parameters are written as
$$
\epsilon=\frac{b^{2}M_{pl}^{2}}{2f_{1}^{2}}\mathrm{coth}^{2}(\phi/f_1), \label{ee1}%
$$
$$
\eta=\frac{bM_{pl}^{2}}{f_{1}^{2}}\left[  (b-1)\mathrm{coth}^{2}(\phi/f_1)+1\right]
$$
and 
$$
\phi=f_1\;\mathrm{cosh}^{-1}\left[  e^{NbM_{pl}^{2}/f^{2}}\mathrm{cosh}%
(\phi_{end}/f_1)\right]  . \label{efold2}%
$$
where $\phi_{end}\simeq\frac{f}{2}\mathrm{ln}\left(  \frac{\theta+1}{\theta
-1}\right)$. Comparing this model with observational data, it has been  found 
$n_{s}\simeq 0.968$, $r\simeq 0.075$, $1<b \le 1.5$ 
and $f_1\ge 11.7M_{pl}$ \cite{Basilakos:2015sza}.
\item Other models of inflation:
The origin of brane \cite{Dvali:2001fw,GarciaBellido:2001ky} which is motivated by the physics of 
extra dimentions  and, on the other hand, the exponential \cite{Goncharov:1985yu,Dvali:1998pa} 
inflationary models are which motivated by the physics of 
extra dimentions. 
It has been found in our study that these models are in agreement
with the Planck data 
although the Starobinsky inflation 
is the winner from the comparison \cite{Ade:2015lrj}.
\end{itemize}
\section{Conclusions}
In this  work we investigated the tachyon inflation on the brane in the context of 
a spatially flat Friedmann-Robertson-Walker  universe. We adopted a specific form of scale factor
from Barrow \cite{Barrow:1996bd} solutions, namely logamediate scale  factor.
Within this context, we estimated analytically the  
slow-roll parameters potential of the model and compare  predictions  
with those of other famous inflationary models in the literature.
Confronting the model against
the latest observational data, we found that the tachyon 
inflationary model on the brane  
is consistent with the results presented in \emph{Planck 2015} within
$1\sigma$ uncertainties for a special class of parameters $(\lambda,A)$ . 
\begin{acknowledgements}:
We would like to thank Ahmad Mehrabi for useful discussions and Mohammad Malekjani for comments on the manuscript and useful discussions. 
\end{acknowledgements}
 \bibliographystyle{apsrev4-1}
  \bibliography{ref}

\begin{thebibliography}{61}%
\makeatletter
\providecommand \@ifxundefined [1]{%
 \@ifx{#1\undefined}
}%
\providecommand \@ifnum [1]{%
 \ifnum #1\expandafter \@firstoftwo
 \else \expandafter \@secondoftwo
 \fi
}%
\providecommand \@ifx [1]{%
 \ifx #1\expandafter \@firstoftwo
 \else \expandafter \@secondoftwo
 \fi
}%
\providecommand \natexlab [1]{#1}%
\providecommand \enquote  [1]{``#1''}%
\providecommand \bibnamefont  [1]{#1}%
\providecommand \bibfnamefont [1]{#1}%
\providecommand \citenamefont [1]{#1}%
\providecommand \href@noop [0]{\@secondoftwo}%
\providecommand \href [0]{\begingroup \@sanitize@url \@href}%
\providecommand \@href[1]{\@@startlink{#1}\@@href}%
\providecommand \@@href[1]{\endgroup#1\@@endlink}%
\providecommand \@sanitize@url [0]{\catcode `\\12\catcode `\$12\catcode
  `\&12\catcode `\#12\catcode `\^12\catcode `\_12\catcode `\%12\relax}%
\providecommand \@@startlink[1]{}%
\providecommand \@@endlink[0]{}%
\providecommand \url  [0]{\begingroup\@sanitize@url \@url }%
\providecommand \@url [1]{\endgroup\@href {#1}{\urlprefix }}%
\providecommand \urlprefix  [0]{URL }%
\providecommand \Eprint [0]{\href }%
\providecommand \doibase [0]{http://dx.doi.org/}%
\providecommand \selectlanguage [0]{\@gobble}%
\providecommand \bibinfo  [0]{\@secondoftwo}%
\providecommand \bibfield  [0]{\@secondoftwo}%
\providecommand \translation [1]{[#1]}%
\providecommand \BibitemOpen [0]{}%
\providecommand \bibitemStop [0]{}%
\providecommand \bibitemNoStop [0]{.\EOS\space}%
\providecommand \EOS [0]{\spacefactor3000\relax}%
\providecommand \BibitemShut  [1]{\csname bibitem#1\endcsname}%
\let\auto@bib@innerbib\@empty
\bibitem [{\citenamefont {Guth}(1981)}]{Guth:1980zm}%
  \BibitemOpen
  \bibfield  {author} {\bibinfo {author} {\bibfnamefont {A.~H.}\ \bibnamefont
  {Guth}},\ }\href {\doibase 10.1103/PhysRevD.23.347} {\bibfield  {journal}
  {\bibinfo  {journal} {Phys. Rev.}\ }\textbf {\bibinfo {volume} {D23}},\
  \bibinfo {pages} {347} (\bibinfo {year} {1981})}\BibitemShut {NoStop}%
\bibitem [{\citenamefont {Albrecht}\ and\ \citenamefont
  {Steinhardt}(1982)}]{Albrecht:1982wi}%
  \BibitemOpen
  \bibfield  {author} {\bibinfo {author} {\bibfnamefont {A.}~\bibnamefont
  {Albrecht}}\ and\ \bibinfo {author} {\bibfnamefont {P.~J.}\ \bibnamefont
  {Steinhardt}},\ }\href {\doibase 10.1103/PhysRevLett.48.1220} {\bibfield
  {journal} {\bibinfo  {journal} {Phys. Rev. Lett.}\ }\textbf {\bibinfo
  {volume} {48}},\ \bibinfo {pages} {1220} (\bibinfo {year}
  {1982})}\BibitemShut {NoStop}%
\bibitem [{\citenamefont {Shtanov}\ \emph {et~al.}(1995)\citenamefont
  {Shtanov}, \citenamefont {Traschen},\ and\ \citenamefont
  {Brandenberger}}]{Shtanov:1994ce}%
  \BibitemOpen
  \bibfield  {author} {\bibinfo {author} {\bibfnamefont {Y.}~\bibnamefont
  {Shtanov}}, \bibinfo {author} {\bibfnamefont {J.~H.}\ \bibnamefont
  {Traschen}}, \ and\ \bibinfo {author} {\bibfnamefont {R.~H.}\ \bibnamefont
  {Brandenberger}},\ }\href {\doibase 10.1103/PhysRevD.51.5438} {\bibfield
  {journal} {\bibinfo  {journal} {Phys. Rev.}\ }\textbf {\bibinfo {volume}
  {D51}},\ \bibinfo {pages} {5438} (\bibinfo {year} {1995})},\ \Eprint
  {http://arxiv.org/abs/hep-ph/9407247} {arXiv:hep-ph/9407247 [hep-ph]}
  \BibitemShut {NoStop}%
\bibitem [{\citenamefont {Kofman}\ \emph {et~al.}(1997)\citenamefont {Kofman},
  \citenamefont {Linde},\ and\ \citenamefont {Starobinsky}}]{Kofman:1997yn}%
  \BibitemOpen
  \bibfield  {author} {\bibinfo {author} {\bibfnamefont {L.}~\bibnamefont
  {Kofman}}, \bibinfo {author} {\bibfnamefont {A.~D.}\ \bibnamefont {Linde}}, \
  and\ \bibinfo {author} {\bibfnamefont {A.~A.}\ \bibnamefont {Starobinsky}},\
  }\href {\doibase 10.1103/PhysRevD.56.3258} {\bibfield  {journal} {\bibinfo
  {journal} {Phys. Rev.}\ }\textbf {\bibinfo {volume} {D56}},\ \bibinfo {pages}
  {3258} (\bibinfo {year} {1997})},\ \Eprint
  {http://arxiv.org/abs/hep-ph/9704452} {arXiv:hep-ph/9704452 [hep-ph]}
  \BibitemShut {NoStop}%
\bibitem [{\citenamefont {Armendariz-Picon}\ \emph {et~al.}(1999)\citenamefont
  {Armendariz-Picon}, \citenamefont {Damour},\ and\ \citenamefont
  {Mukhanov}}]{ArmendarizPicon:1999rj}%
  \BibitemOpen
  \bibfield  {author} {\bibinfo {author} {\bibfnamefont {C.}~\bibnamefont
  {Armendariz-Picon}}, \bibinfo {author} {\bibfnamefont {T.}~\bibnamefont
  {Damour}}, \ and\ \bibinfo {author} {\bibfnamefont {V.~F.}\ \bibnamefont
  {Mukhanov}},\ }\href {\doibase 10.1016/S0370-2693(99)00603-6} {\bibfield
  {journal} {\bibinfo  {journal} {Phys. Lett.}\ }\textbf {\bibinfo {volume}
  {B458}},\ \bibinfo {pages} {209} (\bibinfo {year} {1999})},\ \Eprint
  {http://arxiv.org/abs/hep-th/9904075} {arXiv:hep-th/9904075 [hep-th]}
  \BibitemShut {NoStop}%
\bibitem [{\citenamefont {Sen}(2002{\natexlab{a}})}]{Sen:2002nu}%
  \BibitemOpen
  \bibfield  {author} {\bibinfo {author} {\bibfnamefont {A.}~\bibnamefont
  {Sen}},\ }\href {\doibase 10.1088/1126-6708/2002/04/048} {\bibfield
  {journal} {\bibinfo  {journal} {JHEP}\ }\textbf {\bibinfo {volume} {04}},\
  \bibinfo {pages} {048} (\bibinfo {year} {2002}{\natexlab{a}})},\ \Eprint
  {http://arxiv.org/abs/hep-th/0203211} {arXiv:hep-th/0203211 [hep-th]}
  \BibitemShut {NoStop}%
\bibitem [{\citenamefont {Sen}(2002{\natexlab{b}})}]{Sen:2002an}%
  \BibitemOpen
  \bibfield  {author} {\bibinfo {author} {\bibfnamefont {A.}~\bibnamefont
  {Sen}},\ }\href {\doibase 10.1142/S0217732302008071} {\bibfield  {journal}
  {\bibinfo  {journal} {Mod. Phys. Lett.}\ }\textbf {\bibinfo {volume} {A17}},\
  \bibinfo {pages} {1797} (\bibinfo {year} {2002}{\natexlab{b}})},\ \Eprint
  {http://arxiv.org/abs/hep-th/0204143} {arXiv:hep-th/0204143 [hep-th]}
  \BibitemShut {NoStop}%
\bibitem [{\citenamefont {Sami}\ \emph {et~al.}(2002)\citenamefont {Sami},
  \citenamefont {Chingangbam},\ and\ \citenamefont {Qureshi}}]{Sami:2002fs}%
  \BibitemOpen
  \bibfield  {author} {\bibinfo {author} {\bibfnamefont {M.}~\bibnamefont
  {Sami}}, \bibinfo {author} {\bibfnamefont {P.}~\bibnamefont {Chingangbam}}, \
  and\ \bibinfo {author} {\bibfnamefont {T.}~\bibnamefont {Qureshi}},\ }\href
  {\doibase 10.1103/PhysRevD.66.043530} {\bibfield  {journal} {\bibinfo
  {journal} {Phys. Rev.}\ }\textbf {\bibinfo {volume} {D66}},\ \bibinfo {pages}
  {043530} (\bibinfo {year} {2002})},\ \Eprint
  {http://arxiv.org/abs/hep-th/0205179} {arXiv:hep-th/0205179 [hep-th]}
  \BibitemShut {NoStop}%
\bibitem [{\citenamefont {Kofman}\ and\ \citenamefont
  {Linde}(2002)}]{Kofman:2002rh}%
  \BibitemOpen
  \bibfield  {author} {\bibinfo {author} {\bibfnamefont {L.}~\bibnamefont
  {Kofman}}\ and\ \bibinfo {author} {\bibfnamefont {A.~D.}\ \bibnamefont
  {Linde}},\ }\href {\doibase 10.1088/1126-6708/2002/07/004} {\bibfield
  {journal} {\bibinfo  {journal} {JHEP}\ }\textbf {\bibinfo {volume} {07}},\
  \bibinfo {pages} {004} (\bibinfo {year} {2002})},\ \Eprint
  {http://arxiv.org/abs/hep-th/0205121} {arXiv:hep-th/0205121 [hep-th]}
  \BibitemShut {NoStop}%
\bibitem [{\citenamefont {Setare}\ and\ \citenamefont
  {Kamali}(2012)}]{Setare:2012fg}%
  \BibitemOpen
  \bibfield  {author} {\bibinfo {author} {\bibfnamefont {M.~R.}\ \bibnamefont
  {Setare}}\ and\ \bibinfo {author} {\bibfnamefont {V.}~\bibnamefont
  {Kamali}},\ }\href {\doibase 10.1088/1475-7516/2012/08/034} {\bibfield
  {journal} {\bibinfo  {journal} {JCAP}\ }\textbf {\bibinfo {volume} {1208}},\
  \bibinfo {pages} {034} (\bibinfo {year} {2012})},\ \Eprint
  {http://arxiv.org/abs/1210.0742} {arXiv:1210.0742 [hep-th]} \BibitemShut
  {NoStop}%
\bibitem [{\citenamefont {Setare}\ and\ \citenamefont
  {Kamali}(2013{\natexlab{a}})}]{Setare:2013ula}%
  \BibitemOpen
  \bibfield  {author} {\bibinfo {author} {\bibfnamefont {M.~R.}\ \bibnamefont
  {Setare}}\ and\ \bibinfo {author} {\bibfnamefont {V.}~\bibnamefont
  {Kamali}},\ }\href {\doibase 10.1103/PhysRevD.87.083524} {\bibfield
  {journal} {\bibinfo  {journal} {Phys. Rev.}\ }\textbf {\bibinfo {volume}
  {D87}},\ \bibinfo {pages} {083524} (\bibinfo {year} {2013}{\natexlab{a}})},\
  \Eprint {http://arxiv.org/abs/1305.0740} {arXiv:1305.0740 [hep-th]}
  \BibitemShut {NoStop}%
\bibitem [{\citenamefont {Setare}\ and\ \citenamefont
  {Kamali}(2014{\natexlab{a}})}]{Setare:2014gya}%
  \BibitemOpen
  \bibfield  {author} {\bibinfo {author} {\bibfnamefont {M.~R.}\ \bibnamefont
  {Setare}}\ and\ \bibinfo {author} {\bibfnamefont {V.}~\bibnamefont
  {Kamali}},\ }\href {\doibase 10.1016/j.physletb.2014.10.006} {\bibfield
  {journal} {\bibinfo  {journal} {Phys. Lett.}\ }\textbf {\bibinfo {volume}
  {B739}},\ \bibinfo {pages} {68} (\bibinfo {year} {2014}{\natexlab{a}})},\
  \Eprint {http://arxiv.org/abs/1408.6516} {arXiv:1408.6516 [physics.gen-ph]}
  \BibitemShut {NoStop}%
\bibitem [{\citenamefont {Setare}\ and\ \citenamefont
  {Kamali}(2014{\natexlab{b}})}]{Setare:2014uja}%
  \BibitemOpen
  \bibfield  {author} {\bibinfo {author} {\bibfnamefont {M.~R.}\ \bibnamefont
  {Setare}}\ and\ \bibinfo {author} {\bibfnamefont {V.}~\bibnamefont
  {Kamali}},\ }\href {\doibase 10.1016/j.physletb.2014.07.008} {\bibfield
  {journal} {\bibinfo  {journal} {Phys. Lett.}\ }\textbf {\bibinfo {volume}
  {B736}},\ \bibinfo {pages} {86} (\bibinfo {year} {2014}{\natexlab{b}})},\
  \Eprint {http://arxiv.org/abs/1407.2604} {arXiv:1407.2604 [gr-qc]}
  \BibitemShut {NoStop}%
\bibitem [{\citenamefont {Setare}\ and\ \citenamefont
  {Kamali}(2013{\natexlab{b}})}]{Setare:2013dd}%
  \BibitemOpen
  \bibfield  {author} {\bibinfo {author} {\bibfnamefont {M.~R.}\ \bibnamefont
  {Setare}}\ and\ \bibinfo {author} {\bibfnamefont {V.}~\bibnamefont
  {Kamali}},\ }\href {\doibase 10.1007/JHEP03(2013)066} {\bibfield  {journal}
  {\bibinfo  {journal} {JHEP}\ }\textbf {\bibinfo {volume} {03}},\ \bibinfo
  {pages} {066} (\bibinfo {year} {2013}{\natexlab{b}})},\ \Eprint
  {http://arxiv.org/abs/1302.0493} {arXiv:1302.0493 [hep-th]} \BibitemShut
  {NoStop}%
\bibitem [{\citenamefont {Kamali}\ \emph {et~al.}(2016)\citenamefont {Kamali},
  \citenamefont {Basilakos},\ and\ \citenamefont {Mehrabi}}]{Kamali:2016frd}%
  \BibitemOpen
  \bibfield  {author} {\bibinfo {author} {\bibfnamefont {V.}~\bibnamefont
  {Kamali}}, \bibinfo {author} {\bibfnamefont {S.}~\bibnamefont {Basilakos}}, \
  and\ \bibinfo {author} {\bibfnamefont {A.}~\bibnamefont {Mehrabi}},\ }\href
  {\doibase 10.1140/epjc/s10052-016-4380-6} {\bibfield  {journal} {\bibinfo
  {journal} {Eur. Phys. J.}\ }\textbf {\bibinfo {volume} {C76}},\ \bibinfo
  {pages} {525} (\bibinfo {year} {2016})},\ \Eprint
  {http://arxiv.org/abs/1604.05434} {arXiv:1604.05434 [gr-qc]} \BibitemShut
  {NoStop}%
\bibitem [{\citenamefont {Kamali}\ \emph {et~al.}(2017)\citenamefont {Kamali},
  \citenamefont {Basilakos}, \citenamefont {Mehrabi}, \citenamefont
  {Meysam~Motaharfar},\ and\ \citenamefont {Massaeli}}]{Kamali:2017zgg}%
  \BibitemOpen
  \bibfield  {author} {\bibinfo {author} {\bibfnamefont {V.}~\bibnamefont
  {Kamali}}, \bibinfo {author} {\bibfnamefont {S.}~\bibnamefont {Basilakos}},
  \bibinfo {author} {\bibfnamefont {A.}~\bibnamefont {Mehrabi}}, \bibinfo
  {author} {\bibfnamefont {.}~\bibnamefont {Meysam~Motaharfar}}, \ and\
  \bibinfo {author} {\bibfnamefont {E.}~\bibnamefont {Massaeli}},\ }\href@noop
  {} {\  (\bibinfo {year} {2017})},\ \Eprint {http://arxiv.org/abs/1703.01409}
  {arXiv:1703.01409 [gr-qc]} \BibitemShut {NoStop}%
\bibitem [{\citenamefont {Basilakos}\ \emph {et~al.}(2017)\citenamefont
  {Basilakos}, \citenamefont {Kamali},\ and\ \citenamefont
  {Mehrabi}}]{Basilakos:2017bol}%
  \BibitemOpen
  \bibfield  {author} {\bibinfo {author} {\bibfnamefont {S.}~\bibnamefont
  {Basilakos}}, \bibinfo {author} {\bibfnamefont {V.}~\bibnamefont {Kamali}}, \
  and\ \bibinfo {author} {\bibfnamefont {A.}~\bibnamefont {Mehrabi}},\
  }\href@noop {} {\  (\bibinfo {year} {2017})},\ \Eprint
  {http://arxiv.org/abs/1705.05585} {arXiv:1705.05585 [gr-qc]} \BibitemShut
  {NoStop}%
\bibitem [{\citenamefont {Gibbons}(2002)}]{Gibbons:2002md}%
  \BibitemOpen
  \bibfield  {author} {\bibinfo {author} {\bibfnamefont {G.~W.}\ \bibnamefont
  {Gibbons}},\ }\href {\doibase 10.1016/S0370-2693(02)01881-6} {\bibfield
  {journal} {\bibinfo  {journal} {Phys. Lett.}\ }\textbf {\bibinfo {volume}
  {B537}},\ \bibinfo {pages} {1} (\bibinfo {year} {2002})},\ \Eprint
  {http://arxiv.org/abs/hep-th/0204008} {arXiv:hep-th/0204008 [hep-th]}
  \BibitemShut {NoStop}%
\bibitem [{\citenamefont {Arkani-Hamed}\ \emph {et~al.}(1998)\citenamefont
  {Arkani-Hamed}, \citenamefont {Dimopoulos},\ and\ \citenamefont
  {Dvali}}]{ArkaniHamed:1998rs}%
  \BibitemOpen
  \bibfield  {author} {\bibinfo {author} {\bibfnamefont {N.}~\bibnamefont
  {Arkani-Hamed}}, \bibinfo {author} {\bibfnamefont {S.}~\bibnamefont
  {Dimopoulos}}, \ and\ \bibinfo {author} {\bibfnamefont {G.~R.}\ \bibnamefont
  {Dvali}},\ }\href {\doibase 10.1016/S0370-2693(98)00466-3} {\bibfield
  {journal} {\bibinfo  {journal} {Phys. Lett.}\ }\textbf {\bibinfo {volume}
  {B429}},\ \bibinfo {pages} {263} (\bibinfo {year} {1998})},\ \Eprint
  {http://arxiv.org/abs/hep-ph/9803315} {arXiv:hep-ph/9803315 [hep-ph]}
  \BibitemShut {NoStop}%
\bibitem [{\citenamefont {Arkani-Hamed}\ \emph {et~al.}(1999)\citenamefont
  {Arkani-Hamed}, \citenamefont {Dimopoulos},\ and\ \citenamefont
  {Dvali}}]{ArkaniHamed:1998nn}%
  \BibitemOpen
  \bibfield  {author} {\bibinfo {author} {\bibfnamefont {N.}~\bibnamefont
  {Arkani-Hamed}}, \bibinfo {author} {\bibfnamefont {S.}~\bibnamefont
  {Dimopoulos}}, \ and\ \bibinfo {author} {\bibfnamefont {G.~R.}\ \bibnamefont
  {Dvali}},\ }\href {\doibase 10.1103/PhysRevD.59.086004} {\bibfield  {journal}
  {\bibinfo  {journal} {Phys. Rev.}\ }\textbf {\bibinfo {volume} {D59}},\
  \bibinfo {pages} {086004} (\bibinfo {year} {1999})},\ \Eprint
  {http://arxiv.org/abs/hep-ph/9807344} {arXiv:hep-ph/9807344 [hep-ph]}
  \BibitemShut {NoStop}%
\bibitem [{\citenamefont {Antoniadis}\ \emph {et~al.}(1998)\citenamefont
  {Antoniadis}, \citenamefont {Arkani-Hamed}, \citenamefont {Dimopoulos},\ and\
  \citenamefont {Dvali}}]{Antoniadis:1998ig}%
  \BibitemOpen
  \bibfield  {author} {\bibinfo {author} {\bibfnamefont {I.}~\bibnamefont
  {Antoniadis}}, \bibinfo {author} {\bibfnamefont {N.}~\bibnamefont
  {Arkani-Hamed}}, \bibinfo {author} {\bibfnamefont {S.}~\bibnamefont
  {Dimopoulos}}, \ and\ \bibinfo {author} {\bibfnamefont {G.~R.}\ \bibnamefont
  {Dvali}},\ }\href {\doibase 10.1016/S0370-2693(98)00860-0} {\bibfield
  {journal} {\bibinfo  {journal} {Phys. Lett.}\ }\textbf {\bibinfo {volume}
  {B436}},\ \bibinfo {pages} {257} (\bibinfo {year} {1998})},\ \Eprint
  {http://arxiv.org/abs/hep-ph/9804398} {arXiv:hep-ph/9804398 [hep-ph]}
  \BibitemShut {NoStop}%
\bibitem [{\citenamefont {Binetruy}\ \emph
  {et~al.}(2000{\natexlab{a}})\citenamefont {Binetruy}, \citenamefont
  {Deffayet},\ and\ \citenamefont {Langlois}}]{Binetruy:1999ut}%
  \BibitemOpen
  \bibfield  {author} {\bibinfo {author} {\bibfnamefont {P.}~\bibnamefont
  {Binetruy}}, \bibinfo {author} {\bibfnamefont {C.}~\bibnamefont {Deffayet}},
  \ and\ \bibinfo {author} {\bibfnamefont {D.}~\bibnamefont {Langlois}},\
  }\href {\doibase 10.1016/S0550-3213(99)00696-3} {\bibfield  {journal}
  {\bibinfo  {journal} {Nucl. Phys.}\ }\textbf {\bibinfo {volume} {B565}},\
  \bibinfo {pages} {269} (\bibinfo {year} {2000}{\natexlab{a}})},\ \Eprint
  {http://arxiv.org/abs/hep-th/9905012} {arXiv:hep-th/9905012 [hep-th]}
  \BibitemShut {NoStop}%
\bibitem [{\citenamefont {Binetruy}\ \emph
  {et~al.}(2000{\natexlab{b}})\citenamefont {Binetruy}, \citenamefont
  {Deffayet}, \citenamefont {Ellwanger},\ and\ \citenamefont
  {Langlois}}]{Binetruy:1999hy}%
  \BibitemOpen
  \bibfield  {author} {\bibinfo {author} {\bibfnamefont {P.}~\bibnamefont
  {Binetruy}}, \bibinfo {author} {\bibfnamefont {C.}~\bibnamefont {Deffayet}},
  \bibinfo {author} {\bibfnamefont {U.}~\bibnamefont {Ellwanger}}, \ and\
  \bibinfo {author} {\bibfnamefont {D.}~\bibnamefont {Langlois}},\ }\href
  {\doibase 10.1016/S0370-2693(00)00204-5} {\bibfield  {journal} {\bibinfo
  {journal} {Phys. Lett.}\ }\textbf {\bibinfo {volume} {B477}},\ \bibinfo
  {pages} {285} (\bibinfo {year} {2000}{\natexlab{b}})},\ \Eprint
  {http://arxiv.org/abs/hep-th/9910219} {arXiv:hep-th/9910219 [hep-th]}
  \BibitemShut {NoStop}%
\bibitem [{\citenamefont {Shiromizu}\ \emph {et~al.}(2000)\citenamefont
  {Shiromizu}, \citenamefont {Maeda},\ and\ \citenamefont
  {Sasaki}}]{Shiromizu:1999wj}%
  \BibitemOpen
  \bibfield  {author} {\bibinfo {author} {\bibfnamefont {T.}~\bibnamefont
  {Shiromizu}}, \bibinfo {author} {\bibfnamefont {K.-i.}\ \bibnamefont
  {Maeda}}, \ and\ \bibinfo {author} {\bibfnamefont {M.}~\bibnamefont
  {Sasaki}},\ }\href {\doibase 10.1103/PhysRevD.62.024012} {\bibfield
  {journal} {\bibinfo  {journal} {Phys. Rev.}\ }\textbf {\bibinfo {volume}
  {D62}},\ \bibinfo {pages} {024012} (\bibinfo {year} {2000})},\ \Eprint
  {http://arxiv.org/abs/gr-qc/9910076} {arXiv:gr-qc/9910076 [gr-qc]}
  \BibitemShut {NoStop}%
\bibitem [{\citenamefont {Maartens}\ \emph {et~al.}(2000)\citenamefont
  {Maartens}, \citenamefont {Wands}, \citenamefont {Bassett},\ and\
  \citenamefont {Heard}}]{Maartens:1999hf}%
  \BibitemOpen
  \bibfield  {author} {\bibinfo {author} {\bibfnamefont {R.}~\bibnamefont
  {Maartens}}, \bibinfo {author} {\bibfnamefont {D.}~\bibnamefont {Wands}},
  \bibinfo {author} {\bibfnamefont {B.~A.}\ \bibnamefont {Bassett}}, \ and\
  \bibinfo {author} {\bibfnamefont {I.}~\bibnamefont {Heard}},\ }\href
  {\doibase 10.1103/PhysRevD.62.041301} {\bibfield  {journal} {\bibinfo
  {journal} {Phys. Rev.}\ }\textbf {\bibinfo {volume} {D62}},\ \bibinfo {pages}
  {041301} (\bibinfo {year} {2000})},\ \Eprint
  {http://arxiv.org/abs/hep-ph/9912464} {arXiv:hep-ph/9912464 [hep-ph]}
  \BibitemShut {NoStop}%
\bibitem [{\citenamefont {Cline}\ \emph {et~al.}(1999)\citenamefont {Cline},
  \citenamefont {Grojean},\ and\ \citenamefont {Servant}}]{Cline:1999ts}%
  \BibitemOpen
  \bibfield  {author} {\bibinfo {author} {\bibfnamefont {J.~M.}\ \bibnamefont
  {Cline}}, \bibinfo {author} {\bibfnamefont {C.}~\bibnamefont {Grojean}}, \
  and\ \bibinfo {author} {\bibfnamefont {G.}~\bibnamefont {Servant}},\ }\href
  {\doibase 10.1103/PhysRevLett.83.4245} {\bibfield  {journal} {\bibinfo
  {journal} {Phys. Rev. Lett.}\ }\textbf {\bibinfo {volume} {83}},\ \bibinfo
  {pages} {4245} (\bibinfo {year} {1999})},\ \Eprint
  {http://arxiv.org/abs/hep-ph/9906523} {arXiv:hep-ph/9906523 [hep-ph]}
  \BibitemShut {NoStop}%
\bibitem [{\citenamefont {Csaki}\ \emph {et~al.}(1999)\citenamefont {Csaki},
  \citenamefont {Graesser}, \citenamefont {Kolda},\ and\ \citenamefont
  {Terning}}]{Csaki:1999jh}%
  \BibitemOpen
  \bibfield  {author} {\bibinfo {author} {\bibfnamefont {C.}~\bibnamefont
  {Csaki}}, \bibinfo {author} {\bibfnamefont {M.}~\bibnamefont {Graesser}},
  \bibinfo {author} {\bibfnamefont {C.~F.}\ \bibnamefont {Kolda}}, \ and\
  \bibinfo {author} {\bibfnamefont {J.}~\bibnamefont {Terning}},\ }\href
  {\doibase 10.1016/S0370-2693(99)00896-5} {\bibfield  {journal} {\bibinfo
  {journal} {Phys. Lett.}\ }\textbf {\bibinfo {volume} {B462}},\ \bibinfo
  {pages} {34} (\bibinfo {year} {1999})},\ \Eprint
  {http://arxiv.org/abs/hep-ph/9906513} {arXiv:hep-ph/9906513 [hep-ph]}
  \BibitemShut {NoStop}%
\bibitem [{\citenamefont {Ida}(2000)}]{Ida:1999ui}%
  \BibitemOpen
  \bibfield  {author} {\bibinfo {author} {\bibfnamefont {D.}~\bibnamefont
  {Ida}},\ }\href {\doibase 10.1088/1126-6708/2000/09/014} {\bibfield
  {journal} {\bibinfo  {journal} {JHEP}\ }\textbf {\bibinfo {volume} {09}},\
  \bibinfo {pages} {014} (\bibinfo {year} {2000})},\ \Eprint
  {http://arxiv.org/abs/gr-qc/9912002} {arXiv:gr-qc/9912002 [gr-qc]}
  \BibitemShut {NoStop}%
\bibitem [{\citenamefont {Mohapatra}\ \emph {et~al.}(2000)\citenamefont
  {Mohapatra}, \citenamefont {Perez-Lorenzana},\ and\ \citenamefont
  {de~Sousa~Pires}}]{Mohapatra:2000cm}%
  \BibitemOpen
  \bibfield  {author} {\bibinfo {author} {\bibfnamefont {R.~N.}\ \bibnamefont
  {Mohapatra}}, \bibinfo {author} {\bibfnamefont {A.}~\bibnamefont
  {Perez-Lorenzana}}, \ and\ \bibinfo {author} {\bibfnamefont {C.~A.}\
  \bibnamefont {de~Sousa~Pires}},\ }\href {\doibase 10.1103/PhysRevD.62.105030}
  {\bibfield  {journal} {\bibinfo  {journal} {Phys. Rev.}\ }\textbf {\bibinfo
  {volume} {D62}},\ \bibinfo {pages} {105030} (\bibinfo {year} {2000})},\
  \Eprint {http://arxiv.org/abs/hep-ph/0003089} {arXiv:hep-ph/0003089 [hep-ph]}
  \BibitemShut {NoStop}%
\bibitem [{\citenamefont {Randall}\ and\ \citenamefont
  {Sundrum}(1999)}]{Randall:1999vf}%
  \BibitemOpen
  \bibfield  {author} {\bibinfo {author} {\bibfnamefont {L.}~\bibnamefont
  {Randall}}\ and\ \bibinfo {author} {\bibfnamefont {R.}~\bibnamefont
  {Sundrum}},\ }\href {\doibase 10.1103/PhysRevLett.83.4690} {\bibfield
  {journal} {\bibinfo  {journal} {Phys. Rev. Lett.}\ }\textbf {\bibinfo
  {volume} {83}},\ \bibinfo {pages} {4690} (\bibinfo {year} {1999})},\ \Eprint
  {http://arxiv.org/abs/hep-th/9906064} {arXiv:hep-th/9906064 [hep-th]}
  \BibitemShut {NoStop}%
\bibitem [{\citenamefont {Silverstein}\ and\ \citenamefont
  {Tong}(2004)}]{Silverstein:2003hf}%
  \BibitemOpen
  \bibfield  {author} {\bibinfo {author} {\bibfnamefont {E.}~\bibnamefont
  {Silverstein}}\ and\ \bibinfo {author} {\bibfnamefont {D.}~\bibnamefont
  {Tong}},\ }\href {\doibase 10.1103/PhysRevD.70.103505} {\bibfield  {journal}
  {\bibinfo  {journal} {Phys. Rev.}\ }\textbf {\bibinfo {volume} {D70}},\
  \bibinfo {pages} {103505} (\bibinfo {year} {2004})},\ \Eprint
  {http://arxiv.org/abs/hep-th/0310221} {arXiv:hep-th/0310221 [hep-th]}
  \BibitemShut {NoStop}%
\bibitem [{\citenamefont {Alishahiha}\ \emph {et~al.}(2004)\citenamefont
  {Alishahiha}, \citenamefont {Silverstein},\ and\ \citenamefont
  {Tong}}]{Alishahiha:2004eh}%
  \BibitemOpen
  \bibfield  {author} {\bibinfo {author} {\bibfnamefont {M.}~\bibnamefont
  {Alishahiha}}, \bibinfo {author} {\bibfnamefont {E.}~\bibnamefont
  {Silverstein}}, \ and\ \bibinfo {author} {\bibfnamefont {D.}~\bibnamefont
  {Tong}},\ }\href {\doibase 10.1103/PhysRevD.70.123505} {\bibfield  {journal}
  {\bibinfo  {journal} {Phys. Rev.}\ }\textbf {\bibinfo {volume} {D70}},\
  \bibinfo {pages} {123505} (\bibinfo {year} {2004})},\ \Eprint
  {http://arxiv.org/abs/hep-th/0404084} {arXiv:hep-th/0404084 [hep-th]}
  \BibitemShut {NoStop}%
\bibitem [{\citenamefont {Martin}\ and\ \citenamefont
  {Yamaguchi}(2008)}]{Martin:2008xw}%
  \BibitemOpen
  \bibfield  {author} {\bibinfo {author} {\bibfnamefont {J.}~\bibnamefont
  {Martin}}\ and\ \bibinfo {author} {\bibfnamefont {M.}~\bibnamefont
  {Yamaguchi}},\ }\href {\doibase 10.1103/PhysRevD.77.123508} {\bibfield
  {journal} {\bibinfo  {journal} {Phys. Rev.}\ }\textbf {\bibinfo {volume}
  {D77}},\ \bibinfo {pages} {123508} (\bibinfo {year} {2008})},\ \Eprint
  {http://arxiv.org/abs/0801.3375} {arXiv:0801.3375 [hep-th]} \BibitemShut
  {NoStop}%
\bibitem [{\citenamefont {Guo}\ and\ \citenamefont {Ohta}(2008)}]{Guo:2008sz}%
  \BibitemOpen
  \bibfield  {author} {\bibinfo {author} {\bibfnamefont {Z.-K.}\ \bibnamefont
  {Guo}}\ and\ \bibinfo {author} {\bibfnamefont {N.}~\bibnamefont {Ohta}},\
  }\href {\doibase 10.1088/1475-7516/2008/04/035} {\bibfield  {journal}
  {\bibinfo  {journal} {JCAP}\ }\textbf {\bibinfo {volume} {0804}},\ \bibinfo
  {pages} {035} (\bibinfo {year} {2008})},\ \Eprint
  {http://arxiv.org/abs/0803.1013} {arXiv:0803.1013 [hep-th]} \BibitemShut
  {NoStop}%
\bibitem [{\citenamefont {Kumar}\ \emph {et~al.}(2016)\citenamefont {Kumar},
  \citenamefont {Bueno~Sánchez}, \citenamefont {Escamilla-Rivera},
  \citenamefont {Marto},\ and\ \citenamefont
  {Vargas~Moniz}}]{Escamilla-Rivera:2015ova}%
  \BibitemOpen
  \bibfield  {author} {\bibinfo {author} {\bibfnamefont {K.~S.}\ \bibnamefont
  {Kumar}}, \bibinfo {author} {\bibfnamefont {J.~C.}\ \bibnamefont
  {Bueno~Sánchez}}, \bibinfo {author} {\bibfnamefont {C.}~\bibnamefont
  {Escamilla-Rivera}}, \bibinfo {author} {\bibfnamefont {J.}~\bibnamefont
  {Marto}}, \ and\ \bibinfo {author} {\bibfnamefont {P.}~\bibnamefont
  {Vargas~Moniz}},\ }\href {\doibase 10.1088/1475-7516/2016/02/063} {\bibfield
  {journal} {\bibinfo  {journal} {JCAP}\ }\textbf {\bibinfo {volume} {1602}},\
  \bibinfo {pages} {063} (\bibinfo {year} {2016})},\ \Eprint
  {http://arxiv.org/abs/1504.01348} {arXiv:1504.01348 [astro-ph.CO]}
  \BibitemShut {NoStop}%
\bibitem [{\citenamefont {Ravanpak}\ and\ \citenamefont
  {Salmeh}(2014)}]{Ravanpak:2015bta}%
  \BibitemOpen
  \bibfield  {author} {\bibinfo {author} {\bibfnamefont {A.}~\bibnamefont
  {Ravanpak}}\ and\ \bibinfo {author} {\bibfnamefont {F.}~\bibnamefont
  {Salmeh}},\ }\href {\doibase 10.1103/PhysRevD.89.063504} {\bibfield
  {journal} {\bibinfo  {journal} {Phys. Rev.}\ }\textbf {\bibinfo {volume}
  {D89}},\ \bibinfo {pages} {063504} (\bibinfo {year} {2014})},\ \Eprint
  {http://arxiv.org/abs/1503.06231} {arXiv:1503.06231 [gr-qc]} \BibitemShut
  {NoStop}%
\bibitem [{\citenamefont {Barrow}(1996)}]{Barrow:1996bd}%
  \BibitemOpen
  \bibfield  {author} {\bibinfo {author} {\bibfnamefont {J.~D.}\ \bibnamefont
  {Barrow}},\ }\href {\doibase 10.1088/0264-9381/13/11/012} {\bibfield
  {journal} {\bibinfo  {journal} {Class. Quant. Grav.}\ }\textbf {\bibinfo
  {volume} {13}},\ \bibinfo {pages} {2965} (\bibinfo {year}
  {1996})}\BibitemShut {NoStop}%
\bibitem [{\citenamefont {Barrow}\ and\ \citenamefont
  {Nunes}(2007)}]{Barrow:2007zr}%
  \BibitemOpen
  \bibfield  {author} {\bibinfo {author} {\bibfnamefont {J.~D.}\ \bibnamefont
  {Barrow}}\ and\ \bibinfo {author} {\bibfnamefont {N.~J.}\ \bibnamefont
  {Nunes}},\ }\href {\doibase 10.1103/PhysRevD.76.043501} {\bibfield  {journal}
  {\bibinfo  {journal} {Phys. Rev.}\ }\textbf {\bibinfo {volume} {D76}},\
  \bibinfo {pages} {043501} (\bibinfo {year} {2007})},\ \Eprint
  {http://arxiv.org/abs/0705.4426} {arXiv:0705.4426 [astro-ph]} \BibitemShut
  {NoStop}%
\bibitem [{\citenamefont {Brax}\ and\ \citenamefont {van~de
  Bruck}(2003)}]{Brax:2003fv}%
  \BibitemOpen
  \bibfield  {author} {\bibinfo {author} {\bibfnamefont {P.}~\bibnamefont
  {Brax}}\ and\ \bibinfo {author} {\bibfnamefont {C.}~\bibnamefont {van~de
  Bruck}},\ }\href {\doibase 10.1088/0264-9381/20/9/202} {\bibfield  {journal}
  {\bibinfo  {journal} {Class. Quant. Grav.}\ }\textbf {\bibinfo {volume}
  {20}},\ \bibinfo {pages} {R201} (\bibinfo {year} {2003})},\ \Eprint
  {http://arxiv.org/abs/hep-th/0303095} {arXiv:hep-th/0303095 [hep-th]}
  \BibitemShut {NoStop}%
\bibitem [{\citenamefont {Fairbairn}\ and\ \citenamefont
  {Tytgat}(2002)}]{Fairbairn:2002yp}%
  \BibitemOpen
  \bibfield  {author} {\bibinfo {author} {\bibfnamefont {M.}~\bibnamefont
  {Fairbairn}}\ and\ \bibinfo {author} {\bibfnamefont {M.~H.~G.}\ \bibnamefont
  {Tytgat}},\ }\href {\doibase 10.1016/S0370-2693(02)02638-2} {\bibfield
  {journal} {\bibinfo  {journal} {Phys. Lett.}\ }\textbf {\bibinfo {volume}
  {B546}},\ \bibinfo {pages} {1} (\bibinfo {year} {2002})},\ \Eprint
  {http://arxiv.org/abs/hep-th/0204070} {arXiv:hep-th/0204070 [hep-th]}
  \BibitemShut {NoStop}%
\bibitem [{\citenamefont {Arfken}(1985)}]{Arfken}%
  \BibitemOpen
  \bibfield  {author} {\bibinfo {author} {\bibfnamefont {G.}~\bibnamefont
  {Arfken}},\ }in\ \href@noop {} {\emph {\bibinfo {booktitle} {Mathematical
  Methods for Physicists}}}\ (\bibinfo  {publisher} {FL: Academic Press},\
  \bibinfo {address} {Orlando},\ \bibinfo {year} {1985})\BibitemShut {NoStop}%
\bibitem [{\citenamefont {M.~Abramowitz}(1972)}]{Abramowitz}%
  \BibitemOpen
  \bibfield  {author} {\bibinfo {author} {\bibfnamefont {I.~A.~S.}\
  \bibnamefont {M.~Abramowitz}},\ }in\ \href@noop {} {\emph {\bibinfo
  {booktitle} {Handbook of Mathematical Functions with Formu- las, Graphs, and
  Mathematical Tables, 9th printing}}}\ (\bibinfo  {publisher} {New York:
  Dover},\ \bibinfo {year} {1972})\BibitemShut {NoStop}%
\bibitem [{\citenamefont {Hwang}\ and\ \citenamefont
  {Noh}(2002)}]{Hwang:2002fp}%
  \BibitemOpen
  \bibfield  {author} {\bibinfo {author} {\bibfnamefont {J.-c.}\ \bibnamefont
  {Hwang}}\ and\ \bibinfo {author} {\bibfnamefont {H.}~\bibnamefont {Noh}},\
  }\href {\doibase 10.1103/PhysRevD.66.084009} {\bibfield  {journal} {\bibinfo
  {journal} {Phys. Rev.}\ }\textbf {\bibinfo {volume} {D66}},\ \bibinfo {pages}
  {084009} (\bibinfo {year} {2002})},\ \Eprint
  {http://arxiv.org/abs/hep-th/0206100} {arXiv:hep-th/0206100 [hep-th]}
  \BibitemShut {NoStop}%
\bibitem [{\citenamefont {Langlois}\ \emph {et~al.}(2000)\citenamefont
  {Langlois}, \citenamefont {Maartens},\ and\ \citenamefont
  {Wands}}]{Langlois:2000ns}%
  \BibitemOpen
  \bibfield  {author} {\bibinfo {author} {\bibfnamefont {D.}~\bibnamefont
  {Langlois}}, \bibinfo {author} {\bibfnamefont {R.}~\bibnamefont {Maartens}},
  \ and\ \bibinfo {author} {\bibfnamefont {D.}~\bibnamefont {Wands}},\ }\href
  {\doibase 10.1016/S0370-2693(00)00957-6} {\bibfield  {journal} {\bibinfo
  {journal} {Phys. Lett.}\ }\textbf {\bibinfo {volume} {B489}},\ \bibinfo
  {pages} {259} (\bibinfo {year} {2000})},\ \Eprint
  {http://arxiv.org/abs/hep-th/0006007} {arXiv:hep-th/0006007 [hep-th]}
  \BibitemShut {NoStop}%
\bibitem [{\citenamefont {Ade}\ \emph {et~al.}(2015{\natexlab{a}})\citenamefont
  {Ade} \emph {et~al.}}]{Ade:2015lrj}%
  \BibitemOpen
  \bibfield  {author} {\bibinfo {author} {\bibfnamefont {P.~A.~R.}\
  \bibnamefont {Ade}} \emph {et~al.} (\bibinfo {collaboration} {Planck}),\
  }\href@noop {} {\  (\bibinfo {year} {2015}{\natexlab{a}})},\ \Eprint
  {http://arxiv.org/abs/1502.02114} {arXiv:1502.02114 [astro-ph.CO]}
  \BibitemShut {NoStop}%
\bibitem [{\citenamefont {Ade}\ \emph {et~al.}(2015{\natexlab{b}})\citenamefont
  {Ade} \emph {et~al.}}]{joint}%
  \BibitemOpen
  \bibfield  {author} {\bibinfo {author} {\bibfnamefont {P.}~\bibnamefont
  {Ade}} \emph {et~al.} (\bibinfo {collaboration} {BICEP2, Planck}),\ }\href
  {\doibase 10.1103/PhysRevLett.114.101301} {\bibfield  {journal} {\bibinfo
  {journal} {Phys. Rev. Lett.}\ }\textbf {\bibinfo {volume} {114}},\ \bibinfo
  {pages} {101301} (\bibinfo {year} {2015}{\natexlab{b}})},\ \Eprint
  {http://arxiv.org/abs/1502.00612} {arXiv:1502.00612 [astro-ph.CO]}
  \BibitemShut {NoStop}%
\bibitem [{\citenamefont {Ade}\ \emph {et~al.}(2014)\citenamefont {Ade} \emph
  {et~al.}}]{Planck:2013jfk}%
  \BibitemOpen
  \bibfield  {author} {\bibinfo {author} {\bibfnamefont {P.~A.~R.}\
  \bibnamefont {Ade}} \emph {et~al.} (\bibinfo {collaboration} {Planck}),\
  }\href {\doibase 10.1051/0004-6361/201321569} {\bibfield  {journal} {\bibinfo
   {journal} {Astron. Astrophys.}\ }\textbf {\bibinfo {volume} {571}},\
  \bibinfo {pages} {A22} (\bibinfo {year} {2014})},\ \Eprint
  {http://arxiv.org/abs/1303.5082} {arXiv:1303.5082 [astro-ph.CO]} \BibitemShut
  {NoStop}%
\bibitem [{\citenamefont {Starobinsky}(1980)}]{staro}%
  \BibitemOpen
  \bibfield  {author} {\bibinfo {author} {\bibfnamefont {A.~A.}\ \bibnamefont
  {Starobinsky}},\ }\href {\doibase 10.1016/0370-2693(80)90670-X} {\bibfield
  {journal} {\bibinfo  {journal} {Phys. Lett.}\ }\textbf {\bibinfo {volume}
  {B91}},\ \bibinfo {pages} {99} (\bibinfo {year} {1980})}\BibitemShut
  {NoStop}%
\bibitem [{\citenamefont {Mukhanov}\ and\ \citenamefont
  {Chibisov}(1981)}]{Muk81}%
  \BibitemOpen
  \bibfield  {author} {\bibinfo {author} {\bibfnamefont {V.~F.}\ \bibnamefont
  {Mukhanov}}\ and\ \bibinfo {author} {\bibfnamefont {G.~V.}\ \bibnamefont
  {Chibisov}},\ }\href@noop {} {\bibfield  {journal} {\bibinfo  {journal} {JETP
  Lett.}\ }\textbf {\bibinfo {volume} {33}},\ \bibinfo {pages} {532} (\bibinfo
  {year} {1981})},\ \bibinfo {note} {[Pisma Zh. Eksp. Teor.
  Fiz.33,549(1981)]}\BibitemShut {NoStop}%
\bibitem [{\citenamefont {Ellis}\ \emph {et~al.}(2013)\citenamefont {Ellis},
  \citenamefont {Nanopoulos},\ and\ \citenamefont {Olive}}]{Ellis13}%
  \BibitemOpen
  \bibfield  {author} {\bibinfo {author} {\bibfnamefont {J.}~\bibnamefont
  {Ellis}}, \bibinfo {author} {\bibfnamefont {D.~V.}\ \bibnamefont
  {Nanopoulos}}, \ and\ \bibinfo {author} {\bibfnamefont {K.~A.}\ \bibnamefont
  {Olive}},\ }\href {\doibase 10.1088/1475-7516/2013/10/009} {\bibfield
  {journal} {\bibinfo  {journal} {JCAP}\ }\textbf {\bibinfo {volume} {1310}},\
  \bibinfo {pages} {009} (\bibinfo {year} {2013})},\ \Eprint
  {http://arxiv.org/abs/1307.3537} {arXiv:1307.3537 [hep-th]} \BibitemShut
  {NoStop}%
\bibitem [{\citenamefont {Linde}(1983)}]{Linde}%
  \BibitemOpen
  \bibfield  {author} {\bibinfo {author} {\bibfnamefont {A.~D.}\ \bibnamefont
  {Linde}},\ }\href {\doibase 10.1016/0370-2693(83)90837-7} {\bibfield
  {journal} {\bibinfo  {journal} {Phys. Lett.}\ }\textbf {\bibinfo {volume}
  {B129}},\ \bibinfo {pages} {177} (\bibinfo {year} {1983})}\BibitemShut
  {NoStop}%
\bibitem [{\citenamefont {Barrow}(1990)}]{Barrow:1990vx}%
  \BibitemOpen
  \bibfield  {author} {\bibinfo {author} {\bibfnamefont {J.~D.}\ \bibnamefont
  {Barrow}},\ }\href {\doibase 10.1016/0370-2693(90)90093-L} {\bibfield
  {journal} {\bibinfo  {journal} {Phys. Lett.}\ }\textbf {\bibinfo {volume}
  {B235}},\ \bibinfo {pages} {40} (\bibinfo {year} {1990})}\BibitemShut
  {NoStop}%
\bibitem [{\citenamefont {Barrow}\ and\ \citenamefont
  {Liddle}(1993)}]{Barrow:1993zq}%
  \BibitemOpen
  \bibfield  {author} {\bibinfo {author} {\bibfnamefont {J.~D.}\ \bibnamefont
  {Barrow}}\ and\ \bibinfo {author} {\bibfnamefont {A.~R.}\ \bibnamefont
  {Liddle}},\ }\href {\doibase 10.1103/PhysRevD.47.R5219} {\bibfield  {journal}
  {\bibinfo  {journal} {Phys. Rev.}\ }\textbf {\bibinfo {volume} {D47}},\
  \bibinfo {pages} {5219} (\bibinfo {year} {1993})},\ \Eprint
  {http://arxiv.org/abs/astro-ph/9303011} {arXiv:astro-ph/9303011 [astro-ph]}
  \BibitemShut {NoStop}%
\bibitem [{\citenamefont {Barrow}\ \emph {et~al.}(2006)\citenamefont {Barrow},
  \citenamefont {Liddle},\ and\ \citenamefont {Pahud}}]{Barrow:2006dh}%
  \BibitemOpen
  \bibfield  {author} {\bibinfo {author} {\bibfnamefont {J.~D.}\ \bibnamefont
  {Barrow}}, \bibinfo {author} {\bibfnamefont {A.~R.}\ \bibnamefont {Liddle}},
  \ and\ \bibinfo {author} {\bibfnamefont {C.}~\bibnamefont {Pahud}},\ }\href
  {\doibase 10.1103/PhysRevD.74.127305} {\bibfield  {journal} {\bibinfo
  {journal} {Phys. Rev.}\ }\textbf {\bibinfo {volume} {D74}},\ \bibinfo {pages}
  {127305} (\bibinfo {year} {2006})},\ \Eprint
  {http://arxiv.org/abs/astro-ph/0610807} {arXiv:astro-ph/0610807 [astro-ph]}
  \BibitemShut {NoStop}%
\bibitem [{\citenamefont {Barrow}\ \emph {et~al.}(2014)\citenamefont {Barrow},
  \citenamefont {Lagos},\ and\ \citenamefont {Magueijo}}]{Barrow:2014fsa}%
  \BibitemOpen
  \bibfield  {author} {\bibinfo {author} {\bibfnamefont {J.~D.}\ \bibnamefont
  {Barrow}}, \bibinfo {author} {\bibfnamefont {M.}~\bibnamefont {Lagos}}, \
  and\ \bibinfo {author} {\bibfnamefont {J.}~\bibnamefont {Magueijo}},\ }\href
  {\doibase 10.1103/PhysRevD.89.083525} {\bibfield  {journal} {\bibinfo
  {journal} {Phys. Rev.}\ }\textbf {\bibinfo {volume} {D89}},\ \bibinfo {pages}
  {083525} (\bibinfo {year} {2014})},\ \Eprint {http://arxiv.org/abs/1401.7491}
  {arXiv:1401.7491 [astro-ph.CO]} \BibitemShut {NoStop}%
\bibitem [{\citenamefont {Basilakos}\ and\ \citenamefont
  {Barrow}(2015)}]{Basilakos:2015sza}%
  \BibitemOpen
  \bibfield  {author} {\bibinfo {author} {\bibfnamefont {S.}~\bibnamefont
  {Basilakos}}\ and\ \bibinfo {author} {\bibfnamefont {J.~D.}\ \bibnamefont
  {Barrow}},\ }\href {\doibase 10.1103/PhysRevD.91.103517} {\bibfield
  {journal} {\bibinfo  {journal} {Phys. Rev.}\ }\textbf {\bibinfo {volume}
  {D91}},\ \bibinfo {pages} {103517} (\bibinfo {year} {2015})},\ \Eprint
  {http://arxiv.org/abs/1504.03469} {arXiv:1504.03469 [astro-ph.CO]}
  \BibitemShut {NoStop}%
\bibitem [{\citenamefont {Rubano}\ and\ \citenamefont
  {Barrow}(2001)}]{Rubano:2001xi}%
  \BibitemOpen
  \bibfield  {author} {\bibinfo {author} {\bibfnamefont {C.}~\bibnamefont
  {Rubano}}\ and\ \bibinfo {author} {\bibfnamefont {J.~D.}\ \bibnamefont
  {Barrow}},\ }\href {\doibase 10.1103/PhysRevD.64.127301} {\bibfield
  {journal} {\bibinfo  {journal} {Phys. Rev.}\ }\textbf {\bibinfo {volume}
  {D64}},\ \bibinfo {pages} {127301} (\bibinfo {year} {2001})},\ \Eprint
  {http://arxiv.org/abs/gr-qc/0105037} {arXiv:gr-qc/0105037 [gr-qc]}
  \BibitemShut {NoStop}%
\bibitem [{\citenamefont {Dvali}\ \emph {et~al.}(2001)\citenamefont {Dvali},
  \citenamefont {Shafi},\ and\ \citenamefont {Solganik}}]{Dvali:2001fw}%
  \BibitemOpen
  \bibfield  {author} {\bibinfo {author} {\bibfnamefont {G.~R.}\ \bibnamefont
  {Dvali}}, \bibinfo {author} {\bibfnamefont {Q.}~\bibnamefont {Shafi}}, \ and\
  \bibinfo {author} {\bibfnamefont {S.}~\bibnamefont {Solganik}},\ }in\ \href
  {http://alice.cern.ch/format/showfull?sysnb=2256068} {\emph {\bibinfo
  {booktitle} {{4th European Meeting From the Planck Scale to the Electroweak
  Scale (Planck 2001) La Londe les Maures, Toulon, France, May 11-16, 2001}}}}\
  (\bibinfo {year} {2001})\ \Eprint {http://arxiv.org/abs/hep-th/0105203}
  {arXiv:hep-th/0105203 [hep-th]} \BibitemShut {NoStop}%
\bibitem [{\citenamefont {Garcia-Bellido}\ \emph {et~al.}(2002)\citenamefont
  {Garcia-Bellido}, \citenamefont {Rabadan},\ and\ \citenamefont
  {Zamora}}]{GarciaBellido:2001ky}%
  \BibitemOpen
  \bibfield  {author} {\bibinfo {author} {\bibfnamefont {J.}~\bibnamefont
  {Garcia-Bellido}}, \bibinfo {author} {\bibfnamefont {R.}~\bibnamefont
  {Rabadan}}, \ and\ \bibinfo {author} {\bibfnamefont {F.}~\bibnamefont
  {Zamora}},\ }\href {\doibase 10.1088/1126-6708/2002/01/036} {\bibfield
  {journal} {\bibinfo  {journal} {JHEP}\ }\textbf {\bibinfo {volume} {01}},\
  \bibinfo {pages} {036} (\bibinfo {year} {2002})},\ \Eprint
  {http://arxiv.org/abs/hep-th/0112147} {arXiv:hep-th/0112147 [hep-th]}
  \BibitemShut {NoStop}%
\bibitem [{\citenamefont {Goncharov}\ and\ \citenamefont
  {Linde}(1984)}]{Goncharov:1985yu}%
  \BibitemOpen
  \bibfield  {author} {\bibinfo {author} {\bibfnamefont {A.~S.}\ \bibnamefont
  {Goncharov}}\ and\ \bibinfo {author} {\bibfnamefont {A.~D.}\ \bibnamefont
  {Linde}},\ }\href@noop {} {\bibfield  {journal} {\bibinfo  {journal} {Sov.
  Phys. JETP}\ }\textbf {\bibinfo {volume} {59}},\ \bibinfo {pages} {930}
  (\bibinfo {year} {1984})},\ \bibinfo {note} {[Zh. Eksp. Teor.
  Fiz.86,1594(1984)]}\BibitemShut {NoStop}%
\bibitem [{\citenamefont {Dvali}\ and\ \citenamefont
  {Tye}(1999)}]{Dvali:1998pa}%
  \BibitemOpen
  \bibfield  {author} {\bibinfo {author} {\bibfnamefont {G.~R.}\ \bibnamefont
  {Dvali}}\ and\ \bibinfo {author} {\bibfnamefont {S.~H.~H.}\ \bibnamefont
  {Tye}},\ }\href {\doibase 10.1016/S0370-2693(99)00132-X} {\bibfield
  {journal} {\bibinfo  {journal} {Phys. Lett.}\ }\textbf {\bibinfo {volume}
  {B450}},\ \bibinfo {pages} {72} (\bibinfo {year} {1999})},\ \Eprint
  {http://arxiv.org/abs/hep-ph/9812483} {arXiv:hep-ph/9812483 [hep-ph]}
  \BibitemShut {NoStop}%
\end{thebibliography}%
\end{document}